# Prediction of sustained harmonic walking in the free-living environment using raw accelerometry data

JACEK K. URBANEK [1], VADIM ZIPUNNIKOV [1], TAMARA HARRIS [2],
WILLIAM FADEL [3], NANCY GLYNN [4], ANNEMARIE KOSTER [5],
PAOLO CASEROTTI [6], CIPRIAN CRAINICEANU [1],
JAROSLAW HAREZLAK [7]*

1. Department of Biostatistics, Bloomberg School of Public Health, Johns Hopkins University
2. Laboratory of Epidemiology, Demography, and Biometry, National Institute on Aging
3. Department of Biostatistics, Indiana University School of Medicine
4. Center for Aging and Population Health, Department of Epidemiology, Graduate School of Public Health, University of Pittsburgh
5. Department of Health, Medicine and Life Sciences, Social Medicine, Maastricht University
6. Department of Sports Science and Clinical Biomechanics, University of Southern Denmark
7. Department of Epidemiology and Biostatistics, Indiana University School of Public Health – Bloomington, harezlak@iu.edu

**Objective.** Using raw, sub-second level, accelerometry data, we propose and validate a method for identifying and characterizing walking in the free-living environment. We focus on the sustained harmonic walking (SHW), which we define as walking for at least 10 seconds with low variability of step frequency.
**Approach.** We utilize the harmonic nature of SHW and quantify local periodicity of the tri-axial raw accelerometry data. We also estimate fundamental frequency of observed signals and link it to the instantaneous walking (step-to-step) frequency (IWF). Next, we report total time spent in SHW, number and durations of SHW bouts, time of the day when SHW occurred and IWF for 49 healthy, elderly individuals.
**Main results.** Sensitivity of the proposed classification method was found to be 97%, while specificity ranged between 87% and 97% and prediction accuracy between 94% and 97%. We report total time in SHW between 140 and 10 minutes-per-day distributed between 340 and 50 bouts. We estimate the average IWF to be 1.7 steps-per-second.
**Significance.** We propose a simple approach for detection of SHW and estimation of IWF, based on Fourier decomposition. The resulting approach is fast and allows processing of a week-long raw accelerometry data (approx. 150 million measurements) in relatively short time (~half an hour) on a common laptop computer (2.8 GHz Intel Core i7, 16 GB DDR3 RAM).

Accelerometry, movement recognition, physical activity, walking quantification, wearable computing, free-living data

# 1      Introduction

Accelerometers are now widely used to monitor physical activity in large observational studies where thousands of subjects are observed for weeks at a time. Walking patterns are very often the main focus of such studies as they have been shown to be associated with major health and aging outcomes, including: survival (Studenski *et al* 2011), Parkinson disease (Din *et al* 2016), obesity (Browning 2006) and overall physical capability (Godfrey *et al* 2015). Here, we propose method for estimating when and how people walk based on high-frequency data obtained from wearable accelerometers. These data exhibit extraordinary levels of heterogeneity due to the natural within- and between-person variability, as well as to measuring devices that are prone to batch effects, rotations and random artifacts. Heterogeneity makes walking prediction in the natural environment much more difficult than for "in-the-lab" experiments (Ermes *et al* 2008, Grant *et al* 2006a, Francois *et al* 2009). The differences between data collected "in-the-lab" under strict protocols and data collected in large observational studies is dramatic. Thus made us coin the term data collected "in-the-wild" to emphasize the highly unstructured nature of the data obtained during free-living human activity.

There are a number of accelerometry-based wearable devices designed to recognize and quantify walking "in-the-wild". These include ankle-worn step counters (Coleman *et al* 1999) that provide well-validated information on occurrence and duration of walking as well as the number of steps per minute. Such devices have been widely used in large epidemiological studies, allowing researchers to study walking habits of individuals (Dall *et al* 2015, Orendurff *et al* 2008). Another popular device that has been used to detect walking and posture of individuals is a thigh-worn activPAL monitor (PAL Technologies Ltd, Glasgow, UK). Grant et al. (Grant *et al* 2006) discuss the validity of these devices in posture and activity recognition for "in-the-lab" experiments, whereas in Dall *et al* 2015 and Klenk *et al* 2012 authors use activPAL to collect data on walking patterns in a large group of participants "in-the-wild". While the utility of such devices is undisputable, researchers might often want to address overall physical activity levels rather than the ambulation exclusively (Weuve *et al* 2004). In this paper we propose a method for detection of walking on data collected by hip-worn, accelerometry-based physical activity monitors.

Our work was preceded by a number of approaches. For example papers by Dijkstra *et al* 2008 and Weiss *et al* 2013 both proposed walking-recognition procedure for tri-axial, hip-worn accelerometers. These methods use pre-defined threshold of recorded signal to classify walking activities. The manuscript of Lugade *et al* 2014 introduced an algorithm for classification of postural orientation and movement using data collected by four custom-build activity monitors, while their follow-up work Fortune *et al* 2014 discussed the problem of gait features extraction. The paper by Maurer *et al* 2006 presented a range of classifiers, for different body location of the sensor, including hip-worn devices. Most recently, Hickey *et al* 2017 introduced prediction algorithm dedicated for data collected in free-living environment and validated against video recordings of daily activities.

Definition of walking based on the accelerometer measurements is quite ambiguous. For example, the accelerometry signal has a fundamentally different structure for a 6 minute relaxed walk versus multiple bouts of a few seconds of walking interspersed with other activities. Moreover, laboratory studies typically collect labeled data that is consistent with the 6 minute walking (e.g. 400 meter walking (Chang *et al* 2004) or simply,

6 minute walk task (Enright 2003)). Thus, in practice it is often difficult to validate walking periods that are not consistent with the laboratory-standardized definition without proper gold-standard labels (e.g. video recordings). For these reasons we focus here on sustained harmonic walking (SHW), defined as walking for at least 10 seconds with low variability of step frequency. Based on our observation, we assume that such type of walking is similar in nature to the well-controlled walking, therefore can be identified with methods trained on data collected "in-the-lab" and applied in "in-the-wild" settings.

## 2  Methods
### 2.1  Data collection

The data were collected in the Developmental Epidemiologic Cohort Study (DECOS), a study of healthy older individuals. Participants wore tri-axial ActiGraph GT3X+ device collecting the raw accelerometry signal with a frequency of 80 Hz. The devices were placed on the left hip via an elastic strap. The weight of the devices was 18 grams and the dimensions were 4.6cm x 3.3cm x 1.5cm.

We concentrate on the data from $N = 49$ individuals (24 females and 25 males) who have both "in-the-lab" and "in-the-wild" accelerometry measurements and associated demographic and clinical covariates. Median age of participants was equal to 78.0 years (Q1=74.0, Q3=82.25).

During the "in-the-lab" phase of the experiment all participants were asked to perform a series of physical tasks including 400-meter fast walk, simulated dressing, simulated shopping and chair stands. During 400-meter fast walk participants were instructed to walk 20 laps, 20 meters each, at fast pace. For simulated dressing, participants were instructed to unfold lab jacket, put jacket on (no zipping or buttoning), then remove it, place on a hanger, and put the hanger on a nearby hook. For simulated shopping participants were instructed to walk few steps along the wall, stop by target items, remove them from the upper shelf and place them on the lower shelf. When completed, participants walked back the other direction, moving target items back to the top shelf.

Median time to finish 400-meter walk was 5 minutes 49 seconds (Q1=4 min. 40 s, Q3=6 min.). Dressing activity was set up to last exactly 3 minutes, median time to finish shopping activity was 2 minutes 38 seconds (Q1=1 min. 56 s, Q3=3 min.), median time of standing up from the chair task was 19 seconds (Q1=13 s, Q3=24 s).

"In-the-wild" accelerometry data activities were collected over a one-week period in the natural living environment with no activity labels.

### 2.2  Modeling approach

#### 2.2.1  Definitions and notation

We define SHW as a periodic accelerometry signal performed for at least 10 seconds with a roughly constant walking frequency. Here we propose a method for automatic recognition of SHW intervals together with estimation of the periodicity of the dominating component. The reported frequency of human walking "in-the-wild" ranges between 1.4 - 2.5 Hz (Pachi and Ji 2005). However, we will conservatively focus on the range 1.2 - 4.0 Hz to include slow walking of older individuals and running.

Let $\mathbf{x}(t) = \{x_1(t), x_2(t), x_3(t)\}$ denote the measured signal, where $x_k(t)$ (expressed in g-units) is the measurement at time index $t = 1, \ldots T$ along the axis $k = 1,2,3$. For notational simplicity we drop the subject index. We denote the sampling frequency of the data expressed as number of observations per second by $f_0$.

Our goal is to estimate the walking indicator $y(t)$, where $y(t) = 1$ corresponds to SHW and $y(t) = 0$ corresponds to non-SHW. This is achieved by employing a short-time Fourier transform (STFT) for each axis separately. The discrete Fourier transform (DFT) for axis $k$ is $X_k(f) = \sum_{t=0}^{T-1} x_k(t) e^{-i2\pi f \frac{t}{T}}$, where $f$ is the frequency index, $t$ is the time index and $T$ is the total number of observations. We define the short time Fourier transform (STFT) at time $t$ for axis $k$ of the acceleration signal $\mathbf{x}(t)$ as

$$X_k(t, f; \tau) = \sum_{u=[t-\tau/2]}^{[t+\tau/2]} x_k(u) h(u) e^{-i2\pi f \frac{u}{\tau}}, \tag{1}$$

where $\tau$ is a parameter specifying the number of observations in the interval centered at $t$. We use the Hann window defined as, $h(u; \tau) = 0.5[1 - \cos\{2\pi u/(\tau - 1)\}]$. The spectrum is defined as $S_k(t, f; \tau) = |X_k(t, f; \tau)|$, where $|\cdot|$ denotes the absolute value of a complex number. Next, we introduce the comb function, a function that "combs" the spectrum for those frequencies that are likely to be related to walking (see Figure 1). A comb function is defined by a fundamental frequency $s$, and the thickness of the comb "teeth" with the same width for all frequencies and the total number equal to $n_m$. For any frequency $s$, we define a neighborhood, $N_s = \{s - \frac{1}{U}, s, s + \frac{1}{U}\}$, where $U$ is the duration of the window expressed in seconds, where the local FFT is applied. Thus, $N_s$ is the shortest frequency interval centered at $s$ and consisting of three frequencies. The comb family of functions, $C(f; s)$, indexed by $s$, is defined as

$$C(f; s) = \begin{cases} 1 & \text{for } f \in \cup_{l=2}^{n_m} N_{ls} \\ 0 \end{cases} \tag{2}$$

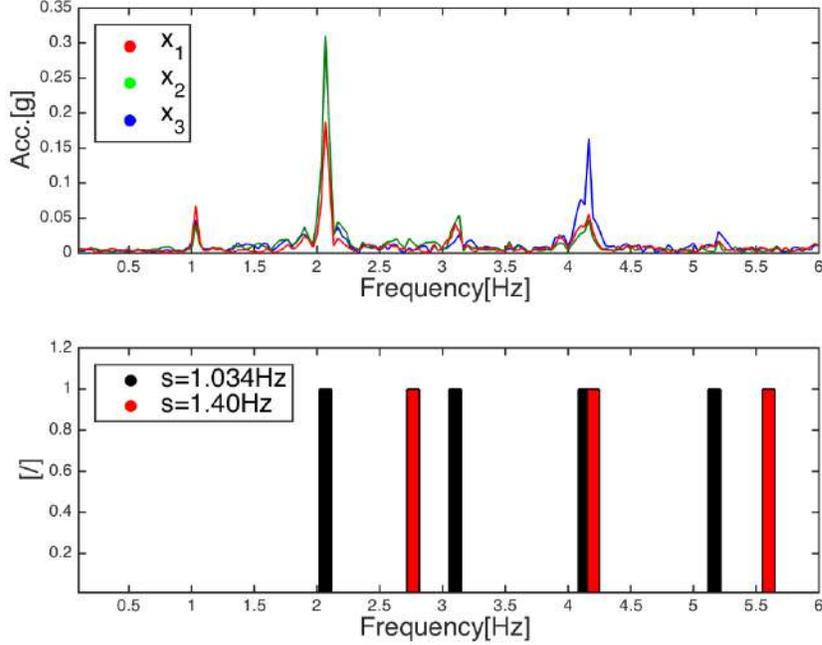

Figure 1: Top figure shows Fourier spectra of tri-axial acceleration signal of walking. Bottom figure shows "Comb" function $C(f;s)$ for $s = w(t) = 1.034$ Hz (black) and for $s = 1.40$ Hz (red). $Black$ lines correspond to the spectral lines representing walking signal components while $red$ lines miss those spectral lines completely.

Spectra shown in the top panel of Figure 1 correspond to a sample walking interval. The bottom panel displays the comb function corresponding to the fundamental frequency of walking acceleration signal, equal to $s = 1.034$ Hz. The tooth centered at $2s = 2.067$ Hz also contains the frequencies between 1.967 and 2.167 Hz. There are additional "comb teeth" centered at $3s = 3.102 Hz$ and so on. Figure 1 also displays an example of another comb shown in red corresponding to $s = 1.40$ Hz, a frequency unrelated to walking acceleration signal. Note that integrating the spectrum in areas corresponding to the "black" comb will result in a higher value than for the spectrum integrating the areas corresponding to the "red" comb, which does not match the spectral peaks. The comb function idea was inspired by the widely used comb filter of harmonic components(Deller et al 2000). However, in contrast to this filter that uses the entire range of spectral components, our comb uses only frequencies between $2s$ and $n_m s$ to limit the number of high and low frequency components and reduce both: high-frequency random noise and low-frequency oscillation resulting from other non-harmonic body movements.

### 2.2.2 Prediction of walking periods and its characteristics

We now provide the technical description of the SHW prediction approach. Specifically, for each axis $k$, we define the area under the full spectrum, $S_k(\cdot)$, as $IS_k(t) = \int_{f=0}^{f_0/2} S_k(f, t; \tau) \, df$ and the area under the spectrum corresponding to the comb function $C(f, s)$ as $IS_k(t, s) = \int_{f=0}^{f_s/2} S_k(f, t; \tau) C(f; s) \, df$. All the functions under the integral are positive and $IS_k(t) > IS_k(t, s)$ for every s and t,. Next, we define

$$Y_k(t,s) = \frac{IS_k(t,s)}{IS_k(t) - IS_k(t,s)} \qquad (3)$$

which is a measure of the size of the periodic content of the accelerometry signal along axis $k$ corresponding to the comb function $C(\cdot;s)$ relative to the total signal along axis $k$.

We use $Y(t,s) = \max_k\{Y_k(t,s)\}$, which is the maximum of the fraction of the signal explained by the frequency $s$ along the three axes $k=1,2,3$ to estimate the SHW periods. For a threshold $\delta$ of $Y(t,s)$, we estimate that a person performs SHW by

$$\hat{y}_\delta(u) = \begin{cases} \text{if } \max_{s \in D_f} Y(t,s) > \delta, \\ 0, \end{cases} \qquad (4)$$

for every $u \in [t - \tau/2, t + \tau/2]$, where $\delta$ is a threshold $\delta$ of $Y(t,s)$, and $D_f$ is the family of frequencies corresponding to walking. We use $D_f = 1.2, \ldots, 4Hz$. The estimation is repeated in 10-second windows shifted by 1 second. The duration of each walking bout is defined as the number of consecutive time windows where walking was estimated to occur plus the window length multiplied by the overlap. For example, if walking was detected in 8 consecutive windows, the walking bout was classified as lasting for 8 + 0.9 x 10 = 17 seconds. It is important to note that we cannot estimate a duration of the walking bout that is shorter than the length of the window $\tau$. Therefore, if walking was detected for only one window it will be classified as a bout lasting for 10 seconds.

For SHW periods, we estimate $IWF$, denoted as $\hat{w}(t)$, as the double of frequency $s$, that maximizes $Y(t,s)$. More precisely,

$$\hat{w}(t) = \begin{cases} 2\, \text{argmax}_{s \in D_f}\{Y(t,s)\} \text{ for } \hat{y}_\delta(t) = 1, \\ Nan, \end{cases} \qquad (5)$$

The complete description of the approach is summarized in Algorithm 1.

---

**Algorithm 1**

**Input:** $\mathbf{x}(t)$ - accelerometry signal, $f_s$ - sampling frequency, $T$ - observation time, $\tau$ - time window, $\delta$ - threshold, $s_{min} = 0.6Hz$, $s_{max} = 2.0Hz$.

**Output:** $y(t)$ - walking indicator, $v(t)$ - vector magnitude, $w(t)$ - IWF.

**Step 1.** Compute the value of the "comb" function $C(f;s)$ for each value of $s \in D_f$.

**Step 2.** Compute the spectrum, $S_k(t,f,\tau)$, for each axis $k = 1,2,3$ and the area under the spectrum $IS_k(t)$ for each $t$.

**Step 3.** Compute the partial area under the spectrum $IS_k(t,s)$ for each $s$ and each $k$.

**Step 4.** Calculate the periodic content of the signal $Y_k(t,s) = \frac{IS_k(t,s)}{IS_k(t)-IS_k(t,s)}$ for each $k$.

**Step 5.** Estimate $y_\delta(t)$ with $\max_{s \in D_f} Y_k(t,s)$.

**Step 6.** For the times $t$ with $\hat{y}_\delta(t) = 1$ estimate $w(t)$ by finding the $s$ that maximizes $Y(t,s)$.

**Step 7.** Identify walking if any of the estimators $\hat{y}_\delta(t)$ is 1.

### 2.3 Validation using "in-the-lab" data

To validate the method, we used data collected during "in-the-lab" phase of the DECOS study.

#### 2.3.1 Selection of the tuning parameters

In this section, we discuss the tuning parameter choices: the window length $\tau$, the threshold $\delta$ of $Y(t,s)$ and the number of harmonics (comb teeth) $n_m$.

There are trade-offs in choosing $\tau$ as: 1) longer windows result in more precise spectrum estimation while 2) shorter time windows are more likely to capture changes in walking frequency or changes in activity type. Based on the empirical evaluation of "in-the-lab" data, we have found that a $\tau = 10$ second interval is long enough for people to maintain their IWF yet short enough to not be sensitive to smooth changes of IWF, resulting in clean spectral signatures.

SHW is estimated in time increments equal to 1 second, so consecutive windows overlap by 90% and a one second interval is declared to be SHW if any of the intervals containing it is estimated to be a SHW. The duration of each walking bout is equal to the number of consecutive time windows where walking was estimated to occur plus the window duration multiplied by the overlap. For example, if walking was detected in 8 consecutive windows, the walking bout was determined to have lasted 8 + 0.9 x 10 = 17 seconds. It is important to note that we cannot estimate a length of the walking bout that is shorter than the length of the window $\tau$. Therefore, if walking was detected for only one window it will be classified as a bout lasting for 10 seconds.

Choice of the threshold $\delta$ and the number of harmonics $n_m$ are inter-related. The proportion of the variability ($\max_{s \in D_f} Y(t,s)$) explained is an increasing function of the number of harmonics $n_m$. We studied $\delta$ as a function of $n_m$, for $n_m = 2, \dots, 17$, where the upper limit 17 is determined by the sampling frequency of the raw accelerometry data. We estimate the density function of $\max_{s \in D_f} Y(t,s)$ for all SHW and non-SHW periods for all subjects. The parameter $\delta$ was then estimated for each subject separately as the intersection between the subject-specific SHW and non-SHW density functions. Figure 2 displays the boxplot of the estimated $\delta$ values across subjects as a function of the number of harmonics, $n_m$. While some between-subject variability exists in the estimation of $\delta$ at every value of of $n_m$, having a population level value simplifies the procedure considerably. For example, when $n_m = 6$ harmonics are used an estimated median value of $\delta$ at the population level is equal to 0.115.

The selection of the number of harmonics, $n_m$, is important for the estimation of IWF, $w(t)$ (equation 5). In principle, we would like to utilize as many frequencies as possible without degrading IWF estimation. To study the choice of $n_m$, we have calculated the IWF, $w(t)$, for every subject at each time point during their 400-meter walk. For every subject, we then calculated the coefficient of variation of IWF during this period. Given that this is a well-controlled experiment, the coefficient of variation of the IWF is expected to be small. Figure 3 displays the coefficient of variation as a function of number of harmonics for every subject. The average coefficient of variation is relatively stable for $n_m = 2, \ldots, 6$ and it starts to increase for larger values of $n_m$. Thus, to use as much periodic information as possible and still keep the coefficient of variation small, we selected $n_m = 6$.

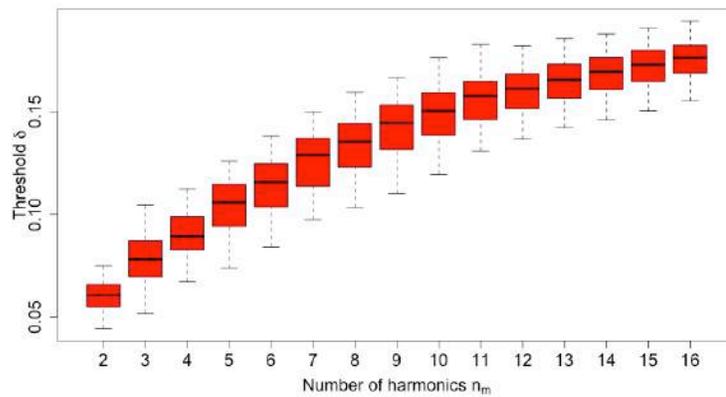

Figure 2: Boxplot representing distributions of subject-specific $\delta$ vs. $n_m$.

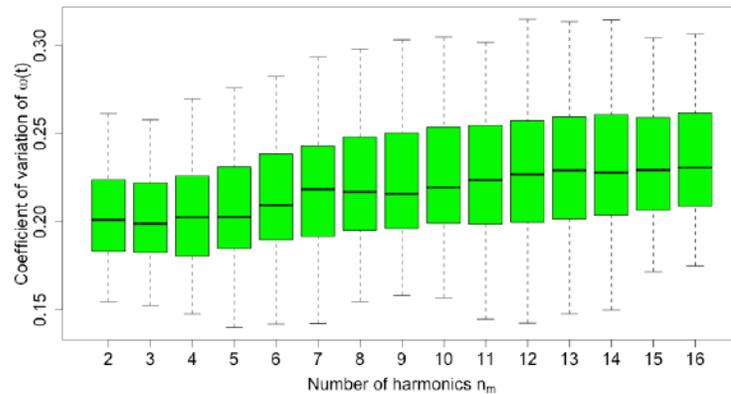

Figure 3: Boxplot representing coefficient of variation as a function of number of harmonics.

### 2.3.2 Performance of the algorithm

For further processing we used $\tau = 10$ and $\delta = 0.115$ and a range of possible IWF

between 1.2 and 4.0 steps/second. The upper boundary for possible IWF was purposefully set to be relatively high to account for running for active individuals.

Next, we calculated specificity, sensitivity and prediction accuracy of the algorithm in three separate classification tasks. Namely, shopping vs. walking, chair stands vs. walking and dressing vs. walking. Shopping, chair stands and dressing were defined as non-walking activities, whereas 400-meter walk was defined as a walking activity. For shopping and chair standing activities specificity was equal to 95% and 97% respectively. Specificity was the lowest for dressing activity and was equal to 87%. Sensitivity of the classifier was equal to 97%. Prediction accuracy was equal to 97% for shopping vs. walking and chair stands vs. walking and 94% for dressing vs. walking.

### 3 Analysis of "in-the-wild" data

We applied the approach described in Section 2.3 to data collected during "in-the-wild" portion of the experiment, to estimate when SHW occurs together with the corresponding IWF. We used the same tuning parameters as in section 2.4.3 ($\tau = 10$ and $\delta = 0.115$ and a range of possible IWF between 1.2 and 4.0 steps/second, with 90% time-window overlap).

Figure 4 displays the total SHW time (top panel) for each of the 49 participants together with the corresponding total number of walking bouts (bottom panel) as per-day averages from the 7 days of activity. Results are sorted in decreasing order according to the estimated average total walking time. Color shading corresponds to different length of walking bouts. For example, the width of the lightest blue bars in the top panel of Figure 4 corresponds to the total walking time from walking bouts equal to 10 seconds. The yellow bars in the bottom panel of Figure 4 display the total number of bouts equal to 10 seconds. Results indicate that long SHW time does not necessarily indicate a large number of SHW bouts. For example, subject 1 had the longest SHW time per day (140 minutes), spread in 210 SHW bouts, which is about the third quartile of the number of SHW bouts for this group. Results indicate that the majority of daily walking bouts for all subjects are between 10 and 30 seconds.

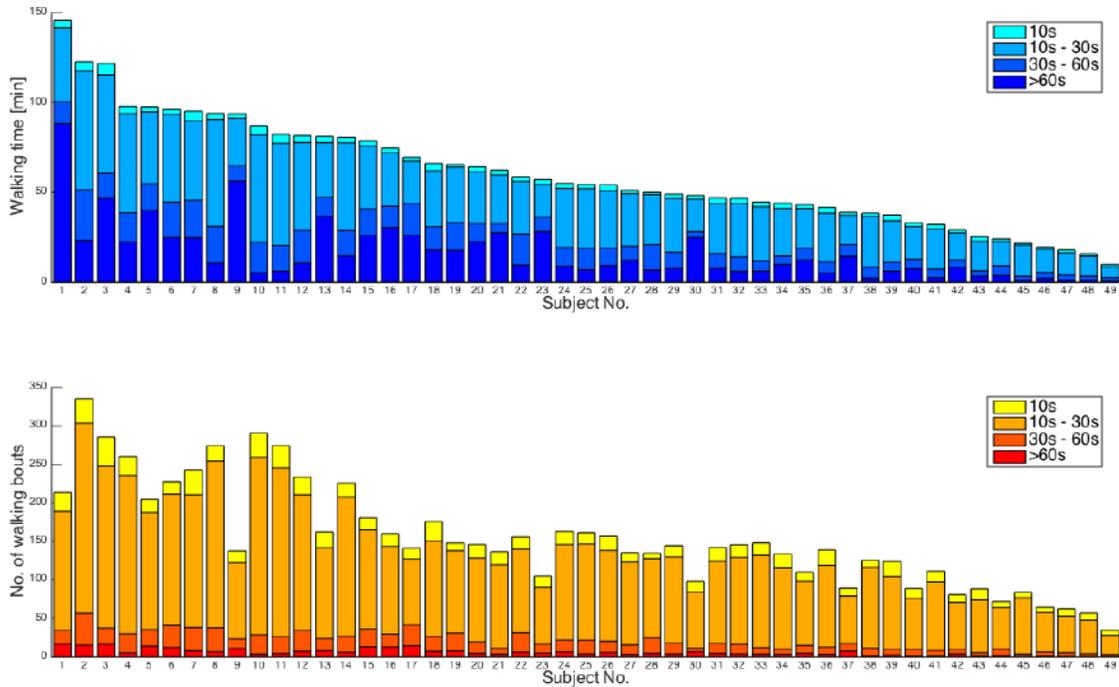
Figure 4: Bar plot presenting the total time of walking and corresponding number of bouts.

Figure 5 provides the lasagna plot (Swihart *et al* 2010) for daily walking characteristics for all subjects. The shades of blue correspond to the number of minutes of walking within a one-hour window. Each row corresponds to one day and dashed red lines separate data for individual subjects. Results were sorted using the same ordering as in Figure 4, with subjects with the highest average daily walking time shown at the top.

Figure 6 displays the boxplots of IWF for each subject sorted according to total walking time. The width of boxes is proportional to the number of walking bouts.

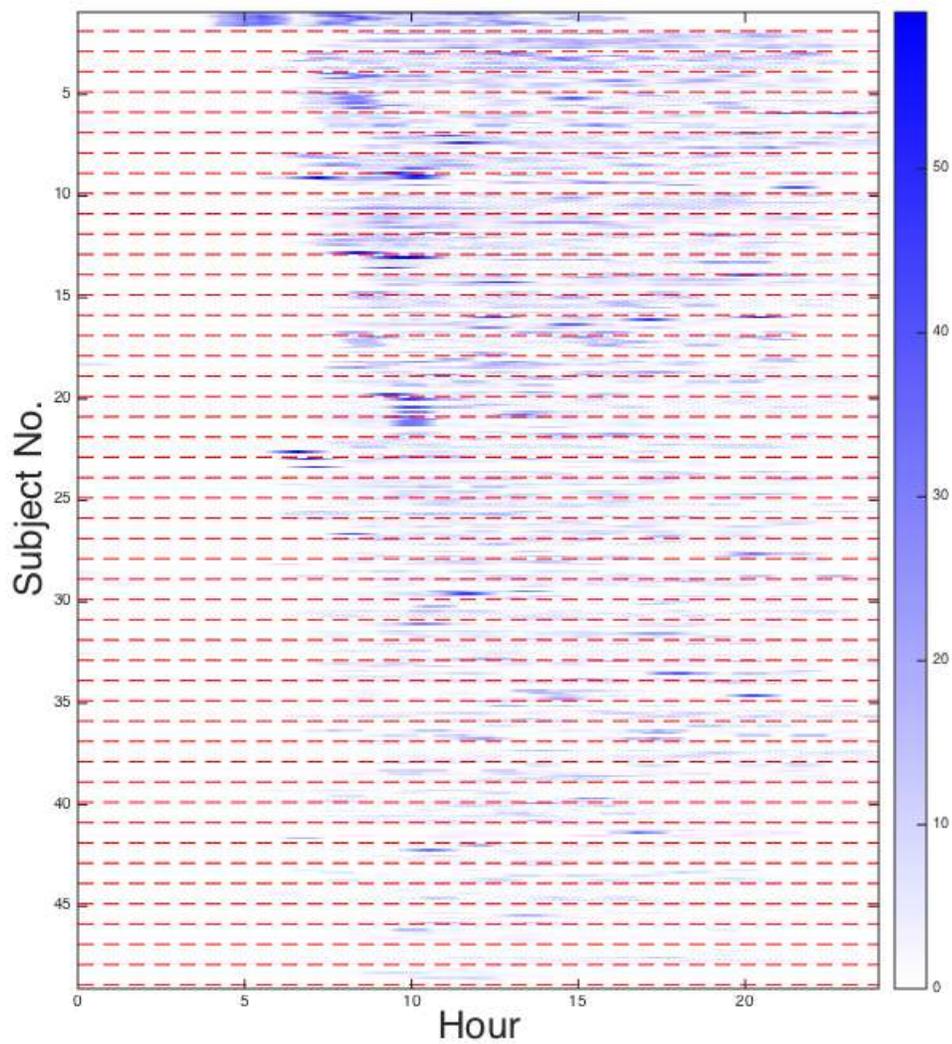

Figure 5: Lasagna plots presenting number of minutes of walking per one hour. Rows represent consecutive days. Red lines separate different subjects.

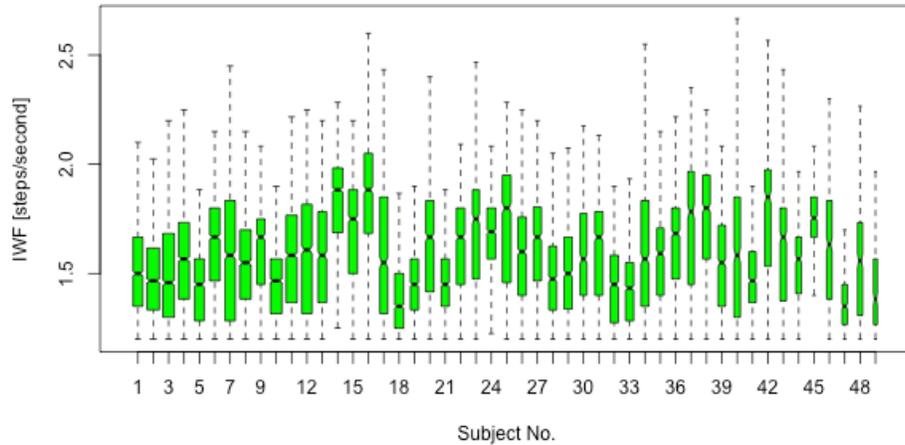

Figure 6: Box-plot presenting IWF for each participant observed "in-the-wild". Width of boxes is proportional to the number of observations.

## 4    Summary and discussion

We have introduced the definition of SHW together with the classification algorithm based on the quantification of the "local degree of periodicity" of all tri-axial time series. The performance of our method depends strongly on the assumption that SHW is a repetitive and sustained process in a particular time window. For example, it is possible for a window of 10 seconds to contain 3 to 4 seconds of walking which are not recognized as walking. Thus, our method will tend to work well at recognizing sustained walking for 10 and more seconds. This can be seen in the results obtained "in-the-lab" where specificity for shopping vs. walking was 95% even though shopping task consists of some non-sustained and harmonic walking. Performance of proposed method is comparable to the previously published results. For example manuscript by Dijkstra *et al* 2008 reported sensitivity of 89.5%, whereas in the paper by Lugade *et al* 2014 authors obtained median sensitivity ranging between 84% to 95% depending on gait velocity. Also, the manuscript by Maurer *et al* 2006 reported prediction accuracy equal to 87% for hip worn accelerometers. That compares favorably to our approach with prediction accuracy between 94% and 97%.

Based on our observations of the data we assumed that SHW is similar in both "in-the-lab" and "in-the-wild" settings and that methods developed using "in-the-lab" data can be scaled-up to the free-living experiment. In reality, however, it is difficult to talk about classifiers performance "in-the-wild" without proper gold standard labels. We believe that our method performs well based on the comparison of obtained results to previously published studies that used accelerometry. For example, paper of Klenk *et al* 2012 reported similar daily walking duration in elderly population, equal to 104.4±50.7 minutes and 102.9±47.8 minutes for males and females respectively. In the manuscript by Orendurff *et al* 2008 authors presented comparable distribution of durations of walking bouts, where 60% of all walking bouts lasted 30 seconds or less, with 20.1% of walking bouts lasting for 10 seconds or less. That suggests significant number of walking bouts that do not fall into a definition of SHW. Reference by Dall *et al* 2015 reported an average value of step frequency around 2.0 steps/second, therefore slightly higher than indicated by our results. Possible explanation of this discrepancy is that the studied population was much younger (mean age 46 years), therefore characterized by higher gait speed (Studenski *et al* 2011) and consequently higher step frequency. Although the arguments given above cannot replace proper validation of our algorithm's performance, they suggest that our method is promising when analyzing high-density accelerometry data collected "in-the-wild".

In future studies, we will explore alternative validation methods for movement recognition "in-the-wild" using multiple wearable sensors as well as investigate the associations between the SHW features and health outcomes.

## 5    Acknowledgements

This research was supported by NIH grant RC2AG036594, Pittsburgh Claude D. Pepper Older Americans Independence Center, Research Registry and Developmental Pilot Grant NIH P30 AG024826 and NIH P30 AG024827 and National Institute on Aging Professional Services Contract HHSN271201100605P. This project was also supported, in part, by the Intramural Research Program of the National Institute on Aging.

This research was supported by the NIH grant RO1 NS085211 from the National Institute of Neurological Disorders and Stroke, by the NIH grant RO1 MH095836 from the National Institute of Mental Health and by NIH grant 1R01 HL123407 from National Hearth, Lung and Blood Institute.

### References

Browning R C 2006 Effects of obesity and sex on the energetic cost and preferred speed of walking *J. Appl. Physiol.* **100** 390–8

Chang M, Cohen-Mansfield J, Ferrucci L, Leveille S, Volpato S, De Rekeneire N and Guralnik J M 2004 Incidence of loss of ability to walk 400 meters in a functionally limited older population *J. Am. Geriatr. Soc.* **52** 2094–2098

Coleman K L, Smith D G, Boone D A, Joseph A W and Aguila M A 1999 Step activity monitor : long-term , continuous recording of ambulatory function **36**

Dall P, Robert P and Mccrorie W 2015 Step Accumulation per Minute Epoch Is Not the Same as Cadence for Free-Living Adults

Deller J R, Proakis J G and Hansen J H L 2000 *Discrete-time processing of speech signals* (IEEE New York, NY, USA:)

Dijkstra B, Zijlstra W, Scherder E, Kamsma Y and Society B G 2008 Detection of walking periods and number of steps in older adults and patients with Parkinson ' s disease : accuracy of a pedometer and an accelerometry-based method **200** 436–441

Din S D, Godfrey A, Galna B, Lord S and Rochester L 2016 Free-living gait characteristics in ageing and Parkinson ' s disease : impact of environment and ambulatory bout

length *J. NeuroEngineering Rehabil.* 1–12

Enright P L 2003 The six-minute walk test *Respir. Care* **48** 783–5

Ermes M, Parkka J, Mantyjarvi J and Korhonen I 2008 Detection of daily activities and sports with wearable sensors in controlled and uncontrolled conditions *Inf. Technol. Biomed. IEEE Trans. On* **12** 20–26

Fortune E, Lugade V, Morrow M and Kaufman K 2014 Validity of using tri-axial accelerometers to measure human movement--Part II: Step counts at a wide range of gait velocities *Med. Eng. Phys.*

Francois S, Chastin M and Granat M 2009 Methods for objective measure , quantification and analysis of sedentary behaviour and inactivity

Godfrey A, Lara J, Munro C A, Wiuff C, Chowdhury S A, Din S D, Hickey A, Mathers J C and Rochester L 2015 Instrumented assessment of test battery for physical capability using an accelerometer : a feasibility study Instrumented assessment of test battery for physical capability using an accelerometer : a feasibility study

Grant P M, Ryan C G, Tigbe W W and Granat M H 2006 The validation of a novel activity monitor in the measurement of posture and motion during everyday activities *Br J Sports Med* **40** 992–997

Hickey A, Del Din S, Rochester L and Godfrey A 2017 Detecting free-living steps and walking bouts: validating an algorithm for macro gait analysis *Physiol. Meas.* **38** N1–15

Klenk J, Bu G, Rapp K, Franke S and Peter R 2012 Walking on sunshine : effect of weather conditions on physical activity in older people 2010–3

Lugade V, Fortune E, Morrow M and Kaufman K 2014 Validity of Using Tri-Axial Accelerometers to Measure Human Movement - Part I: Posture and Movement Detection *Med Eng Phys* **36** 169–76

Maurer U, Smailagic A, Siewiorek D P and Deisher M 2006 Activity recognition and monitoring using multiple sensors on different body positions *Wearable and Implantable Body Sensor Networks, 2006. BSN 2006. International Workshop on* (IEEE) p 4–pp

Orendurff M S, Schoen J A, Bernatz G C, Segal A D and Klute G K 2008 How humans walk : Bout duration , steps per bout , and rest duration **45**

Pachi A and Ji T 2005 Frequency and velocity of people walking *Struct. Eng.* **83**

Studenski S, Perera S, Patel K, Rosano C, Faulkner K, Inzitari M, Brach J, Chandler J, Cawthon P, Connor E B and Others 2011 Gait speed and survival in older adults *Jama* **305** 50–Swihart B J, Caffo B, James B D, Strand M, Schwartz B S and Punjabi N M 2010 Lasagna plots: a saucy alternative to spaghetti plots *Epidemiol. Camb. Mass* **21** 621

Weiss A, Brozgol M, Dorfman M, Herman T, Shema S, Giladi N and Hausdorff J M 2013 Does the evaluation of gait quality during daily life provide insight into fall risk? A novel approach using 3-day accelerometer recordings *Neurorehabil. Neural Repair* **27** 742–752

Weuve J, Kang J H, Manson J E, Breteler M M B, Ware J H and Grodstein F 2004 Physical Activity , Including Walking , and Cognitive Function in Older Women **292**